\documentclass[anz]{cupaus}

\usepackage{mathtools}
\usepackage{appendix}
\usepackage{nicefrac} 

\begin{document}

\verso{Fully  3D  Rayleigh-Taylor Instability in a Boussinesq Fluid}
\recto{Fully  3D  Rayleigh-Taylor Instability in a Boussinesq Fluid}

\title{Fully  3D  Rayleigh-Taylor Instability in a Boussinesq Fluid}

\cauthormark 
\author[1]{S.J. Walters}

\author[2]{L.K. Forbes}

\address[1]{School of Mathematics and Physics, University of Tasmania, P.O. Box 37, Hobart, 7001, Tasmania, Australia\email[1]{stephen.walters@utas.edu.au}
ORCID: 0000-0002-4480-767X}

\address[2]{School of Mathematics and Physics, University of Tasmania, P.O. Box 37, Hobart, 7001, Tasmania, Australia\email[1]{larry.forbes@utas.edu.au}
ORCID: 0000-0002-9135-3594}
      
\pages{1}{6}
      
\begin{abstract}
Rayleigh-Taylor instability occurs when a heavier fluid overlies a lighter fluid, and the two seek to exchange positions under the effect of gravity.  We present linearized theory for arbitrary 3D initial disturbances that grow in time, and calculate the evolution of the interface for early times.  A new spectral method is introduced for the fully 3D non-linear problem in a Boussinesq fluid, where the interface between the light and heavy fluid is approximated with a smooth but rapid density change in the fluid.  The results of large-scale numerical calculation are presented in fully 3D geometry, compared and contrasted with the early-time linearized theory.
\end{abstract}

\keywords[2010 \textit{Mathematics subject classification}]{76E17: Interfacial stability and instability; 76M22: Spectral methods}

\keywords[\textit{Keywords and phrases}]{Viscous flow, Rayleigh-Taylor instability, linearization, spectral method}

\maketitle

\section{Introduction}
\label{sec1}
The accurate modelling of the flow of fluids is an area of ongoing interest and intense research. Such studies have been central in the development of numerous modern technological advances including, but certainly not limited to, heavier-than-air flight, aero-dynamics and hydro-dynamics, aeronautics, supersonic flight, weather prediction. The growth of instabilities is an important area of study, as it leads to an understanding of regions of significant change in the system, such as energy dissipation in the case of vortex formation.

In particular, in this paper we examine the Rayleigh-Taylor instability, which describes the gravitationally unstable situation where a denser fluid sits upon a less dense fluid. Any small disturbance to the initially horizontal interface will grow, allowing the denser fluid to move down and the lighter fluid to rise. This problem was originally described by Rayleigh \cite{ray} and further studied by Taylor \cite{tay}. Since that time, much work has been carried out investigating behaviour, largely of the two dimensional case. Increasingly, however, modelling of fully three dimensional systems is becoming possible with the ongoing increases in computing performance. The review by Sharp \cite{sharp} gives a summary of some of the work undertaken in this large field of research. More recently, reviews by Abarzhi \cite{abar} and Zhou (\cite{zhou1} and \cite{zhou2}) particularly consider the energy dissipation aspect and the self-similar nature of Rayleigh-Taylor mixing, and hence the transition to turbulent flow. Apart from finite difference, finite element, and spectral methods, modern approaches to the numerical simulation of the Rayleigh-Taylor instability include the use of the Cahn-Hilliard equation with various numerical schemes (see \cite{lee}), the Lattice Boltzmann method \cite{liang} and smoothed particle hydrodynamics methods \cite{shadloo}, to mention only a few of the large number of approaches to this problem.

The purpose of this paper is to study three-dimensional (3D) Rayleigh-Taylor flow, and determine to what extent linearization might retain some predictive power at early times, and to explore some fully 3D outflow morphologies, from simple initial disturbances. To do this we will use the Boussinesq approximation (see \cite{farrow}), which is a valid approach if the densities of the two fluids are similar. The velocity-vorticity formulation will obviate the need to solve for the pressure explicitly, which is computationally expensive to do accurately. A bi-stream function approach will avoid the need to solve the incompressibility equation (Laplace's equation for the velocity potential) at each time step, so eliminating another expensive computational task. These stream functions and the density perturbation will be represented as Fourier series, which means that derivatives can be evaluated explicitly in terms of the basis functions. Such spectral methods exhibit exponential convergence (\cite{boyd}, pp. 45-46) and are eminently suitable for solutions which are smooth and in simple geometries, both of which apply to the problem studied in this paper. Finally, the resultant large set of independent calculations will be solved on a Graphical Processing Unit (GPU) which is designed to handle many simple calculations in parallel. We will demonstrate that this combination of approaches leads to a highly accurate modelling of fully 3D fluid flow problems using enough precision to show clearly the features of the flow.

\section{Theoretical Framework}
\label{sec2}
The equations describing the Rayleigh-Taylor problem for incompressible fluids are the equation for conservation of mass (incompressibility condition), equation (\ref{mass}) and the incompressible Navier-Stokes equation (\ref{ns}):
\begin{eqnarray}
\label{mass}
\boldsymbol \nabla \cdot \boldsymbol q&=&0\\
\label{ns}
\frac{\partial \boldsymbol q}{\partial t}+(\boldsymbol q \cdot \boldsymbol \nabla)\boldsymbol q-\frac{\mu}{\rho}\nabla^2 \boldsymbol q&=&\boldsymbol g-\frac{\boldsymbol \nabla p}{\rho}.
\end{eqnarray}
Here, $\boldsymbol q$ is the velocity field, $\mu$ is the dynamic viscosity, $\rho$ is the density, $p$ is the pressure and $\boldsymbol g$ is the acceleration due to gravity.

Whether using primitive variables or spectral or other methods, these equations are computationally expensive to solve numerically, usually requiring the forward integration of a large system of differential equations.

In the current paper we use a 3D spectral method similar to that used by Forbes \cite{forbes2009} who solved the viscous Rayleigh-Taylor problem in two dimensions using a Boussinesq approximation.

In recent years, computational speeds have continued to increase, and this paper re-examines the problem with more powerful computer hardware. We have been able to extend Forbes' original work into a fully 3D problem. A decrease in required computing time of more than two orders of magnitude has been achieved with newer hardware and parallelisation. This has allowed us to compute the evolution of the interface in three dimensions with sufficient accuracy and precision to observe in detail the expected roll-over of the interface. In addition, as the problem can now be made fully 3D, we will present images of flow resulting from initial conditions which are neither spherically nor axially symmetric.

\section{Linearized Inviscid Model}
\label{sec3}
We first consider the simpler case where we assume that there is no viscosity in the two fluids. This will give a solution to the Rayleigh-Taylor instability which will be valid for early times. Such a solution provides a means of validating the numerical solution, at least for early times, and gives confidence that the numerical scheme is suitable. The following linearisation is similar to that of \cite{tay}, but is slightly more general in that it includes the effects due to finite depth, contains a 3D analysis, rather than two-dimensional (2D), and also discusses the difference in behaviour due to different modes in the initial disturbance.

In this inviscid case, if the two layers of fluid are initially at rest, and the interface between them is a small perturbation to a known function, a linearised system can be constructed which is valid for early times. In the 3D Rayleigh-Taylor model, we assume that the interface is a horizontal plane with a small perturbation function.

The problem is constructed such that the denser fluid layer `fluid 2' sits upon the lighter layer of `fluid 1'. We specify that the `floor' is at $z=-H_1$ and the `ceiling' is at $z=H_2$. The initial interface between the fluids is situated at $z=0$. The assumption that the fluids are inviscid means that the flow is irrotational, $\nabla \times \boldsymbol q = 0$. This condition is met through the use of a velocity potential for each fluid, so that in each fluid the velocity field $ \boldsymbol q$ takes the form $ \boldsymbol q=\nabla \Phi$ for a potential field $\Phi$.  This allows integration of the equations of motion for each fluid ($j=1,2$) to give the Bernoulli equations (\cite{batch}, p. 383).

\begin{eqnarray}
\frac{\partial \Phi_j}{\partial t}+\frac{1}{2}(u_j^2+v_j^2+w_j^2)+\frac{p_j}{\rho_j}+g z=0 \hspace{1cm}, \hspace{1cm} j=1,2 .
\label{bernoulli}
\end{eqnarray}

In addition, since the flow is irrotational in this inviscid case, the substitution of $\boldsymbol q=\nabla \Phi$ into the conservation of mass equations (\ref{mass}), results in Laplace's equation in each fluid
\begin{eqnarray}
\nabla^2 \Phi_j&=&0  \hspace{1cm}, \hspace{1cm} j=1,2. \nonumber 
\end{eqnarray}
The functions $\Phi_1, \Phi_2$ are the velocity potentials in the two fluids.

We consider an interface between the two fluids at $z=\eta(x,y,t)$. The pressure immediately to either side of the interface must be equal, so that at $z=\eta$, $p_1=p_2$. Rearranging equation (\ref{bernoulli}) for pressure, this equality at the interface gives the dynamic constraint

\begin{eqnarray}
\rho_1\bigg[\frac{\partial \Phi_1}{\partial t}+\frac{1}{2}(u_1^2+v_1^2+w_1^2)+g z\bigg]=\rho_2\bigg[\frac{\partial \Phi_2}{\partial t}+\frac{1}{2}(u_2^2+v_2^2+w_2^2)+g z\bigg]
\label{interface1}
\end{eqnarray}
on $z=\eta$.

We now introduce dimensionless variables, by scaling with the gravitational constant $g$, and the size of the computational window $L$. The conversion to these variables is:

\begin{eqnarray}
(\hat{x},\hat{y},\hat{z},\hat{H_1},\hat{H_2},\hat{\eta})&=& \frac{2\pi}{L}(x,y,z,H_1,H_2,\eta)\nonumber\\
\hat t&=&\sqrt{\frac{2\pi g}{L}}t\nonumber\\
\hat {\boldsymbol q}&=&\sqrt{\frac{2 \pi}{g L}}\boldsymbol q\nonumber\\
\hat{\Phi}&=&\sqrt{\frac{8\pi^3}{g L^3}}\Phi\nonumber.
\end{eqnarray}
We will use these new variables, but drop the hats for readability. In these variables, and introducing $D=\rho_2/\rho_1$ as the dimensionless ratio of densities, equation (\ref{interface1}) becomes

\begin{eqnarray}
\frac{\partial \Phi_1}{\partial t}+\frac{1}{2}(u_1^2+v_1^2+w_1^2)+z=D\bigg[\frac{\partial \Phi_2}{\partial t}+\frac{1}{2}(u_2^2+v_2^2+w_2^2)+z\bigg] \nonumber
\end{eqnarray}
on $z=\eta$.

In the state where the interface is perfectly flat at $z=0$, and where both fluids are at rest, equation (\ref{interface1}) is satisfied, as all the terms in the brackets are zero. We will expand about this zero-state (unstable) equilibrium using a small dimensionless parameter $\epsilon$ for the shape of the interface and the velocity potentials:
\begin{eqnarray}
\eta(x,y,t) \equiv 0+\epsilon \eta^L(x,y,t)+O(\epsilon^2)\nonumber\\
\Phi_1(x,y,z,t) \equiv 0+\epsilon \Phi_1^L(x,y,z,t)+O(\epsilon^2)\nonumber\\
\Phi_2(x,y,z,t) \equiv 0+\epsilon \Phi_2^L(x,y,z,t)+O(\epsilon^2)\nonumber
\end{eqnarray}
where the superscript L denotes the linearised variable. 

The linearised problem then takes the form of Laplace's equation for each fluid:
\begin{eqnarray}
\nabla^2 \Phi_2^L&=&0 \quad\quad \text{in}\quad 0<z<H_2\nonumber\\
\label{laplace2}
\nabla^2 \Phi_1^L&=&0 \quad\quad \text{in}\quad -H_1<z<0
\end{eqnarray}
where the kinematic boundary conditions are:
\begin{eqnarray}
\frac{\partial{\Phi_1^L}}{\partial z}&=&0 \quad\quad \text{on}\quad z=-H_1\nonumber\\
\frac{\partial{\Phi_2^L}}{\partial z}&=&0 \quad\quad \text{on}\quad z=H_2\nonumber\\
\label{phi1}
\frac{\partial{\Phi_1^L}}{\partial z}&=&\frac{\partial{\eta^L}}{\partial t} \quad\quad \text{on}\quad z=0\\
\frac{\partial{\Phi_2^L}}{\partial z}&=&\frac{\partial{\eta^L}}{\partial t}\quad\quad \text{on}\quad z=0. \nonumber
\end{eqnarray}
The dynamic boundary condition, equation (\ref{interface1}) linearizes to:
\begin{eqnarray}\label{dyn}
(D-1)\eta^L+D\frac{\partial{\Phi_2^L}}{\partial t}-\frac{\partial{\Phi_1^L}}{\partial t}&=&0 \quad\quad \text{on}\quad z=0
\end{eqnarray}

We begin formulating a solution to Laplace's equation (\ref{laplace2}) in each fluid separately. Using standard separation of variables techniques, for an $(m,n)$-th Fourier mode, we may write

\begin{eqnarray}
\label{phi1L}
\Phi_1^L(x,y,z,t)=A_1(t)\cos(mx)\cos(ny)\cosh(\xi(z+H_1))\\
\Phi_2^L(x,y,z,t)=A_2(t)\cos(mx)\cos(ny)\cosh(\xi(H_2-z))\nonumber
\end{eqnarray} 
where we have written $\xi\equiv\sqrt{m^2+n^2}$ for convenience. Since the problem is linear, we may later superpose these models for a more general shape of the initial interface.

We now need to solve for $A_1(t)$ and $A_2(t)$. Describing the $(m,n)$-th mode of the interface as $\eta^L(x,y,t)=G(t)\cos(mx)\cos(ny)$, and substituting into equation (\ref{phi1}), as well as equation (\ref{phi1L}), we obtain
\begin{eqnarray}
\label{dhdt}
\frac{\text d G(t)}{\text d t}&=&A_1(t)\xi\sinh(\xi H_1)
\end{eqnarray}
Then equation (\ref{dyn}), which is the dynamic condition at $z=0$, may be written as
\begin{eqnarray}
(D-1)G(t)+D\frac{\text d A_2}{\text dt}\cosh(\xi H_2)-\frac{\text d A_1}{\text dt}\cosh(\xi H_1)=0 \nonumber
\end{eqnarray}
With equation (\ref{dhdt}), this yields a second order linear equation for $G(t)$ which may be solved to yield:
\begin{eqnarray}
G(t)=C_1 e^{\alpha t}+C_2 e^{-\alpha t} \nonumber
\end{eqnarray}
where the constant $\alpha$ is given by
\begin{eqnarray}
\label{alpha}
\alpha=\sqrt{\frac{(D-1)\xi}{D\coth(\xi H_2)+\coth(\xi H_1)}}.
\end{eqnarray}
This parameter $\alpha$ depends on integers $m,n$ and gives the growth rate for the $(m,n)$-th mode. The constants $C_1$ and $C_2$ are determined by the initial conditions. If, for example, the interface starts from rest, then $G'(t)=0$ at $t=0$, so that $C_1=C_2$ and then $G(t)= C_{mn}\cosh(\alpha t)$ for an interface disturbance at the $(m,n)$-th mode. Thus the linearised behaviour of any interface may be written as the sum of contributions from each mode:
\begin{eqnarray}
\eta(x,y,t)=\epsilon\sum_{m=0}^{\infty} \sum_{n=0}^{\infty}C_{mn}\cosh(\alpha t)\cos(mx)\cos(ny)+O(\epsilon^2).
\label{linear1}
\end{eqnarray}
The $C_{mn}$ coefficients are determined from the initial shape of the interface. So the linearised solution predicts that any small disturbance to the initially flat interface will grow exponentially, with growth rates given by (\ref{alpha}). The highest modes have largest $m,n$ and therefore largest $\alpha$ and so they will dominate. That is, the linearised solution at time $t$ will be a superposition of the initial modes multiplied by exponential factors. As expected this solution cannot `roll over', but does give an accurate description of the behaviour of the interface for early times. We will compare this with the numerical solution to the non-linear viscous problem in Section \ref{sec6}.

\section{Boussinesq Viscous Model}
\label{sec4}
We now include viscosity in the system, using a Boussinesq approximation. As in Section \ref{sec3}, we are considering a system with a denser fluid sitting on top of a lighter fluid. However, in the Boussinesq model \cite{farrow}, we approximate this by a single fluid with a density which changes rapidly but smoothly across the region where the sharp interface would be in the two fluids model. This difference in density will drive the motion as it did in the linear model of Section \ref{sec3}. In this non-linear model, however, we note that the density difference will slowly be reduced by diffusion over time.

Returning briefly to dimensional variables, let the original two fluids have densities $\rho_1$ and $\rho_2$. Then we define the single density for the Boussinesq fluid as a constant (the original density of fluid 1) and a small perturbation to that density which varies over time and space: $\rho(x,y,z,t)=\rho_1+\tilde{\rho}(x,y,z,t)$. 
The equations governing this system are now the incompressibility condition, the transport equation for the density perturbation, and the incompressible Navier-Stokes-Boussinesq equation:
\begin{eqnarray}
\boldsymbol \nabla \cdot\boldsymbol q = 0 \nonumber\\
\frac{\partial \tilde \rho}{\partial t}+(\boldsymbol q \cdot \boldsymbol \nabla) \tilde \rho = \sigma \nabla^2 \tilde \rho \nonumber\\
\label{i-ns}
\frac{\partial \boldsymbol q}{\partial t}+(\boldsymbol q \cdot \boldsymbol \nabla) \boldsymbol q+\frac{1}{\rho_1}\boldsymbol\nabla p = \frac{1}{\rho_1}\big[-(\rho_1+\tilde \rho)g \boldsymbol k+\mu \nabla^2 \boldsymbol q \big]
\end{eqnarray}
Here $\mu$ is the dynamic viscosity, $\boldsymbol{k}$ is the unit vector in the $z$-direction, and $\sigma$ is the diffusion coefficient for density. It is related to a Prandtl number, as discussed by Farrow and Hocking \cite{farrow}. In order to avoid calculating the pressure in this last equation, which is computationally expensive, and not necessary for determing the flow, we will solve the vorticity equation instead. By taking the curl of equation (\ref{i-ns}), we remove the pressure term from the equation. Defining the vorticity vector field $\boldsymbol \zeta \equiv \{\zeta^X,\zeta^Y,\zeta^Z\}=\boldsymbol{\nabla} \times \boldsymbol{q}$, the curl of the momentum equation gives the vorticity equation (\cite{batch}, p.267):
\begin{eqnarray}
\frac{\partial \boldsymbol \zeta}{\partial t}+(\boldsymbol q \cdot \boldsymbol \nabla)\boldsymbol \zeta-(\boldsymbol{\zeta} \cdot \boldsymbol{\nabla})\boldsymbol{q}=-\frac{g}{\rho_1} \boldsymbol\nabla \times (\tilde\rho\boldsymbol k)+\mu\nabla^2\boldsymbol\zeta.
\label{vort1}
\end{eqnarray}

As with the linearised inviscid problem, we now switch to dimensionless variables using the same conversion constants $g$, $L$  and $\rho_1$.  The conversion for the two additional variables is
\begin{eqnarray}
\hat{\tilde \rho}&=&\frac{\tilde \rho}{\rho_1}\nonumber\\
\hat{\boldsymbol \zeta}&=&\sqrt{\frac{L}{2\pi g}}\boldsymbol\zeta\nonumber.
\end{eqnarray}

In dimensionless variables equations (\ref{i-ns}) and (\ref{vort1}) then become:
\begin{eqnarray}
\boldsymbol{\hat{\nabla}} \cdot\boldsymbol{\hat{q}} = 0 \nonumber\\
\frac{\partial \hat{\tilde \rho}}{\partial t}+(\boldsymbol{\hat{q}} \cdot {\hat{\boldsymbol \nabla}}) \hat{\tilde \rho} = \hat{\sigma} \hat{\nabla}^2 \hat{\tilde \rho} \label{transport}\\
\frac{\partial \boldsymbol{\hat{\zeta}}}{\partial \hat{t}}+(\boldsymbol{\hat{q}} \cdot \boldsymbol{\hat{\nabla}})\boldsymbol{\hat{\zeta}}-(\boldsymbol{\hat{\zeta}} \cdot \boldsymbol{\hat{\nabla}})\boldsymbol{\hat{q}}=-\boldsymbol{\hat{\nabla}} \times (\tilde{\hat{\rho}}\boldsymbol k)+\hat{\mu}\hat{\nabla}^2\boldsymbol{\hat{\zeta}}.
\label{i-ns2}
\end{eqnarray}
where $\hat{\mu}=\mu(2\pi/L)^{3/2}g^{-1/2}\rho_1^{-1}$. Notice that the dimensionless viscosity $\hat{\mu}$ here is the inverse of a Reynolds number. As in section \ref{sec3}, we will use these dimensionless variables, but for readability we will omit the hats.

In order to confine our computations to a finite region, we stipulate that fluid may not cross the boundary of the computational window. Thus we require the following conditions on the velocity components:
\begin{eqnarray}
u&=&0 \text{ at } x=\pm \pi\nonumber\\
v&=&0 \text{ at } y=\pm \pi\nonumber\\
w&=&0 \text{ at } z=-H_1,H_2
\label{boundaries}
\end{eqnarray}

The incompressibility condition and these boundary conditions are met using a certain spectral representation. The incompressibility condition is satisfied identically by choosing to let the velocity field $\boldsymbol q$ be the curl of some vector field $\boldsymbol V \equiv \{V_1,V_2,V_3\}$. Only two independent components are required, so we are free to choose one of these components. Following the approach of Forbes and Brideson \cite{forbes2017}, we let $V_3=0$ and rename $V_1\equiv \Psi,V_2\equiv \chi $. Using lower case coordinate subscripts to denote partial derivatives, the velocity components are:
\begin{eqnarray}
u&=&-\chi_z\nonumber\\
v&=&\Psi_z\nonumber\\
w&=&\chi_x-\Psi_y\nonumber
\end{eqnarray}
and the incompressibility condition
\begin{eqnarray}
u_x+v_y+w_z=0 \nonumber
\end{eqnarray}
is met identically.
In order to meet the boundary conditions (\ref{boundaries}), we use a spectral representation of $\Psi,\chi$ as
\begin{eqnarray}
\chi(x,y,z,t)=\sum_{m=0}^{\infty} \sum_{n=0}^{\infty} \sum_{p=1}^{\infty} A_{mnp}(t) \sin(\alpha_m x)\cos(\alpha_n y)\sin(\alpha_p (z+H_1))\nonumber\\
\Psi(x,y,z,t)=\sum_{m=0}^{\infty} \sum_{n=0}^{\infty} \sum_{p=1}^{\infty} B_{mnp}(t) \cos(\alpha_m x)\sin(\alpha_n y)\sin(\alpha_p (z+H_1))
\label{spectral1}
\end{eqnarray}
with the coefficients $\alpha_m,\alpha_n, \alpha_p$ chosen to meet the boundary conditions, and the integers $m,n,p$ are the spectral modes. We initially chose basis functions with $\alpha_m=m,\alpha_n=n, \alpha_p=p \pi/(H_1+H_2)$. This means that the particular solution in this case will be left-right symmetric in $x$ and $y$, but need not be symmetric in $z$. If we wish to examine solutions which are not necessarily symmetric about the $x$ and $y$ axes, we can choose basis functions as we have for $z$. We will also  use a spectral representation for the density term $\tilde\rho$:
\begin{eqnarray}
\tilde\rho(x,y,z,t) = \sum_{m=0}^{\infty} \sum_{n=0}^{\infty} \sum_{p=1}^{\infty} C_{mnp}(t) \cos(\alpha_m x)\cos(\alpha_n y)\sin(\alpha_p (z+H_1))
\label{spectral2}
\end{eqnarray}

With the choice of sine for the $z$ basis functions, this forces the fluid to have zero density perturbation on the top and bottom boundary at all times, as can be seen in Fig. \ref{fig1}. To correct this restriction, simulations were then run with $\alpha_p=p \pi /(H_1+H_2)/2$, which allows the density to take any value throughout the fluid. This change may be seen in Fig. \ref{fig2} and Fig. \ref{fig3}, where there are no thin layers of fluid at top or bottom of the region.
We substitute these representations into the equations for vorticity (\ref{i-ns2}) and transport (\ref{transport}), and solve for the sets of derivatives of coefficients $A_{mnp},B_{mnp}$ and $C_{mnp}$. Each coefficient can be isolated by multiplication by the appropriate basis functions and integrating over the computational space, using the orthogonality of the trigonometric functions. The final system of equations to be solved numerically is too lengthy to include here, but is described in the appendix (\ref{appendix}).

\section{Notes on Computing System and Performance}
\label{sec5}
The 3D computations reported in this paper were performed using an Intel I7-7700 CPU with 32 GB RAM and a Quadro GP100 GPU running Ubuntu 17.10. The computations were written in Fortran and compiled using the PGI Fortran compiler \cite{pgi}, with ACC-directives moving most of the calculations to the GPU. The output from the program was an array of data for $\tilde\rho$ at each chosen time. This data was read into \textsc{Matlab} to produce the figures.

\begin{figure}[ht]
\includegraphics[width=\columnwidth,clip]{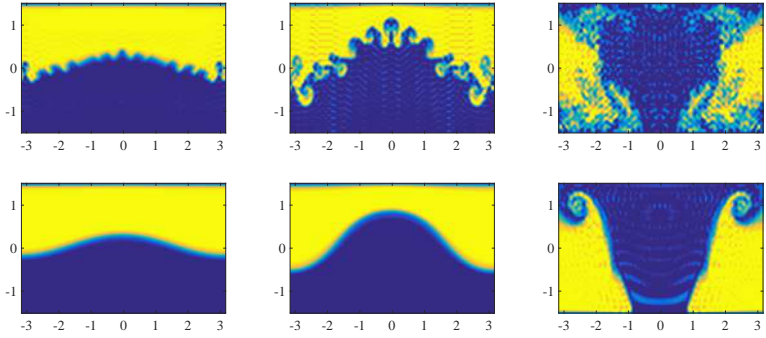}
\caption{A 2D slice of the 3D interface is shown at dimensionless times $t=21$ (left), $t=28$ (centre), $t=42$ (right) for calculations performed using single precision (upper) and double precision (lower). The use of single precision values in the upper images has produced multiple numerical artefacts which obscure the solution more and more as time passes. The use of double precision values removed these artefacts as seen in the lower three images. Colour available online.}
  \label{fig1}
\end{figure}

In the process of testing our novel spectral techniques, it was discovered that numerous unexpected artefacts became apparent in the 3D models. This was determined to be due to the use of single precision calculations, for which many graphics cards are optimised. This problem was resolved by using a GPU designed for fast double precision calculations. In Figure \ref{fig1}, we show the difference between single precision calculation and double precision in an otherwise identical numerical routine. The top three panels illustrate the growth of a single-mode Rayleigh-Taylor instability, with density ratio $D=1.05$, from initial amplitude $\epsilon = 0.03$, at the three times $t=21, 28, 42$, computed at single precision. The lower three panels are for the identical conditions, except that here the results were generated using double precision arithmetic.

Both sets of results were taken from the centre plane $y=0$ of a fully 3D calculation. Clearly the single precision results in the upper three frames have suffered the growth of inaccuracies due to the use of 32 bit floats in the computations and the resultant accumulation of round off errors. This is equivalent to adding small disturbances in very high modes. Once these higher modes are formed, they then grow as Rayleigh-Taylor instabilities in their own right. As shown in the linearised analysis, higher modes grow faster than lower modes, at least initially.

No such problems occur with the more accurate double precision results shown in the lower three diagrams at the same three times, where now the interfacial roll-up at the last time $t=42$ is clearly visible. Although computer run times were increased by a factor of two to three times, it was nevertheless found to be necessary to run all 3D programs using double precision arithmetic, due to the extremely large number of computations needed in fully 3D computations.

\section{Results}
\label{sec6}
\subsection{Two-dimensional calculations}
\label{2dcalc}
\begin{figure}[ht]
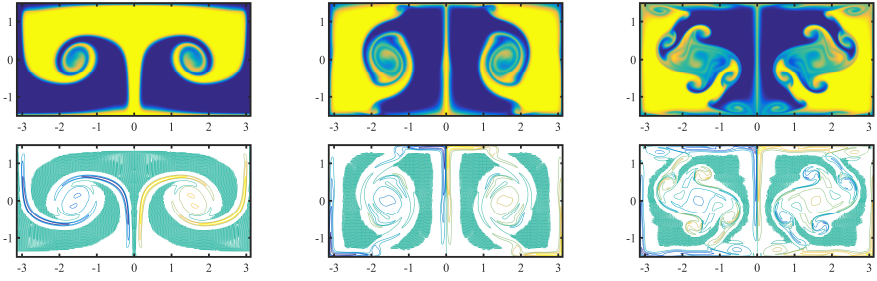

  \begin{center}
  \includegraphics[width=\columnwidth/22*7]{t45_640_196_ntim400.eps}
  \includegraphics[width=\columnwidth/22*7]{t54_640_196_ntim400.eps}  
  \includegraphics[width=\columnwidth/22*7]{t63_640_196_ntim400.eps}\\
  \includegraphics[width=\columnwidth/22*7]{vrt_zeta45_640_196_ntim400.eps}
  \includegraphics[width=\columnwidth/22*7]{vrt_zeta54_640_196_ntim400.eps}  
  \includegraphics[width=\columnwidth/22*7]{vrt_zeta63_640_196_ntim400.eps}
  \caption{2D interface, with initial interface perturbation shown at times $t=45$ (left column), $t=54$ (middle column) and $t=63$ (right column). The less dense fluid is shown in a lighter shade (yellow online) and the denser fluid in a darker shade (blue online). These calculations used 1280 points and 382 modes in the $x$-direction and 640 points and 191 modes in $y$. The two axes have equal scales. The upper row shows the density, while the lower row shows the vorticity.}
   \label{fig2}
  \end{center}
 
\end{figure}

Repeating the earlier 2D model of Forbes \cite{forbes2009} with the approach described in Section \ref{sec5} resulted in a decrease in computational time by a factor of approximately 250. This allowed the use of a much greater number of mesh points and spectral modes, which improved accuracy, allowing the integration to continue far past the `roll-over' point. The results are also able to be observed with much higher resolution. In Figure \ref{fig2} we present the rolled up 2D interface at dimensionless times $t=45$, $t=54$ and $t=63$. The surface started from an initial interface shape $y=\epsilon \cos x$, with $\epsilon=0.03$. Specifically the initial density perturbation is set at $\tilde\rho=0.05/(1 + Exp(100 (0.03 \cos(x) - y)))$. This gives an interfacial region in which the density perturbation changes smoothly, but very rapidly from $\tilde\rho=0$ to $\tilde\rho=0.05$. The system of spectral coefficients described in Section \ref{sec4} in equations (\ref{spectral1}) and (\ref{spectral2}), and detailed in the appendix (but limited to two dimensions), were incorporated into the transport and momentum equations (\ref{i-ns}), and were integrated forward in time using the classical fourth order Runge-Kutta method given in Atkinson (\cite{atkinson}, p.371). This can be adapted easily for 2D and 3D systems of ordinary differential equations, which allows parallelization of the computation. The less dense fluid (yellow online) initially has dimensionless density parameter $1$, and the denser fluid (blue online) has density ratio $D=1.05$. The computational window extends from $-\pi$ to $\pi$ in the $x$-direction, and from $-1.5$ to $1.5$ in the $z$-direction. This last constraint causes the typical mushroom shape to flatten out as it approaches the ceiling. If a less constrained flow is desired, the computational `walls' can be moved further away, or the initial perturbation can be reduced in width (the latter approach was taken in the 3-D case below). We see that as the lighter fluid rises through the denser fluid, it entrains the denser fluid, causing it both to roll up and over and then to rise again in two central plumes. The run-time for this computation at 1280 by 640 grid points, and 382 by 191 spectral modes was 14 hours.  Past the times shown, the fluids continue to swirl around, and gradually their densities equalise through diffusion. These Kelvin-Helmholtz type roll-up effects are characteristic of Rayleigh-Taylor instabilities where the density ratio is small. Daly \cite{daly} describes the variation in this effect at density ratios of $1.1, 2$ and $10$. While spiral formation is found at $D=1.1$, none is seen in the high density difference $D=10$ case. The Boussinesq approximation adopted in this paper assumes that the density ratio is close to $1$, and we set $D=1.05$ in all simulations.

\begin{figure}[ht]
  \begin{center}
  \includegraphics[width=\columnwidth/22*7]{t45.eps}
  \includegraphics[width=\columnwidth/22*7]{t60.eps}  
  \includegraphics[width=\columnwidth/22*7]{t75.eps}\\
  \includegraphics[width=\columnwidth/22*7]{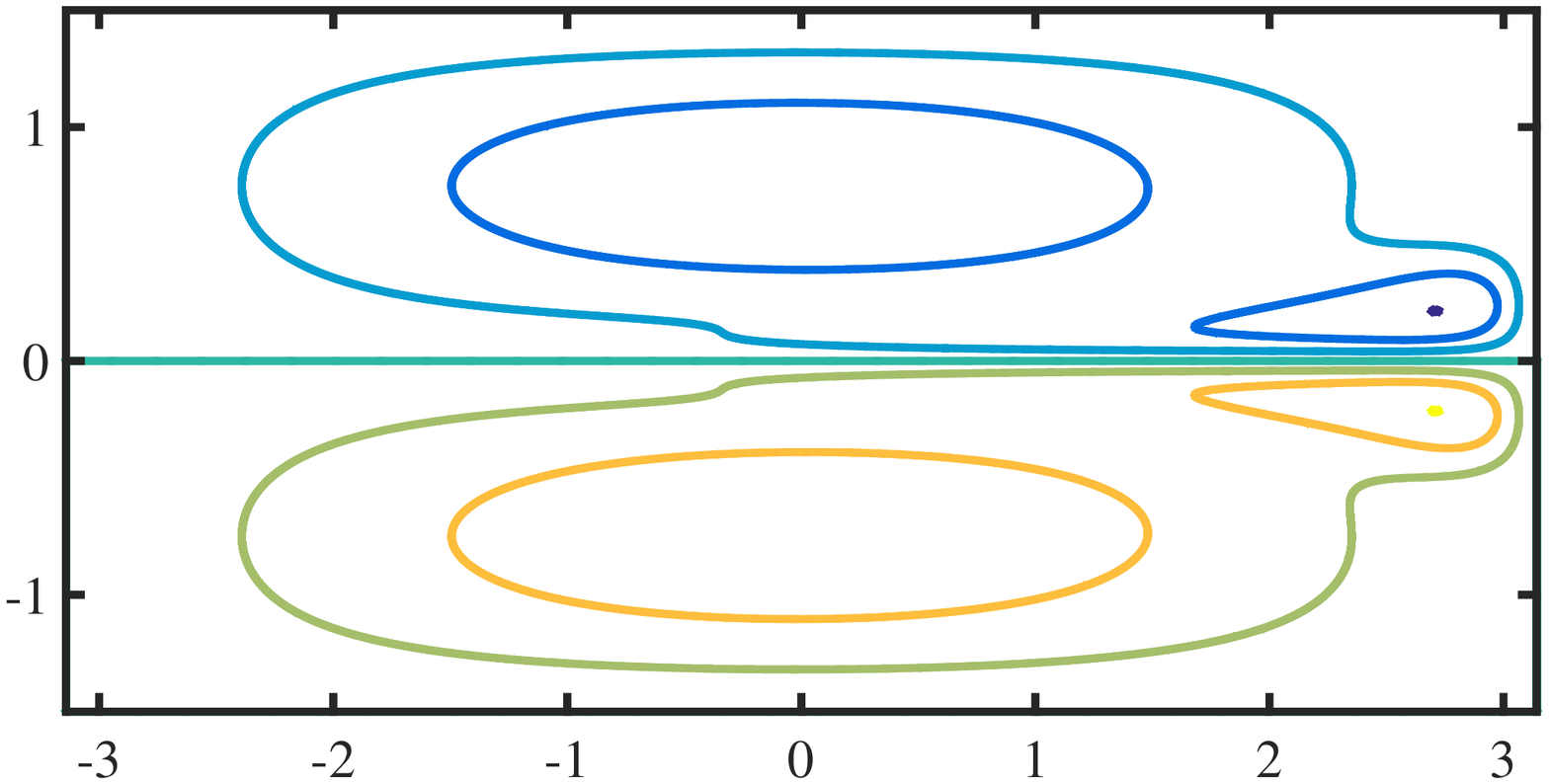}
  \includegraphics[width=\columnwidth/22*7]{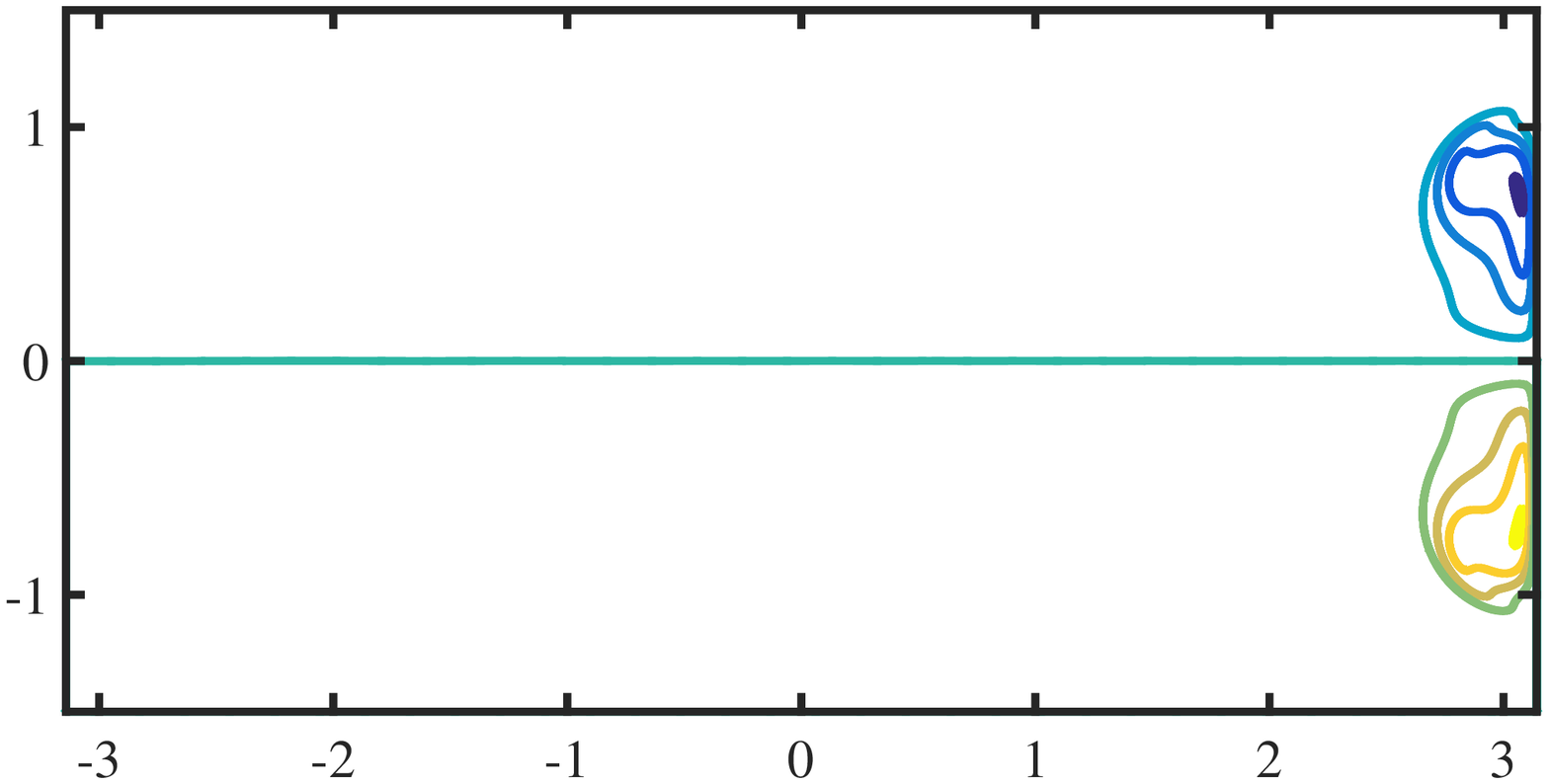}
  \includegraphics[width=\columnwidth/22*7]{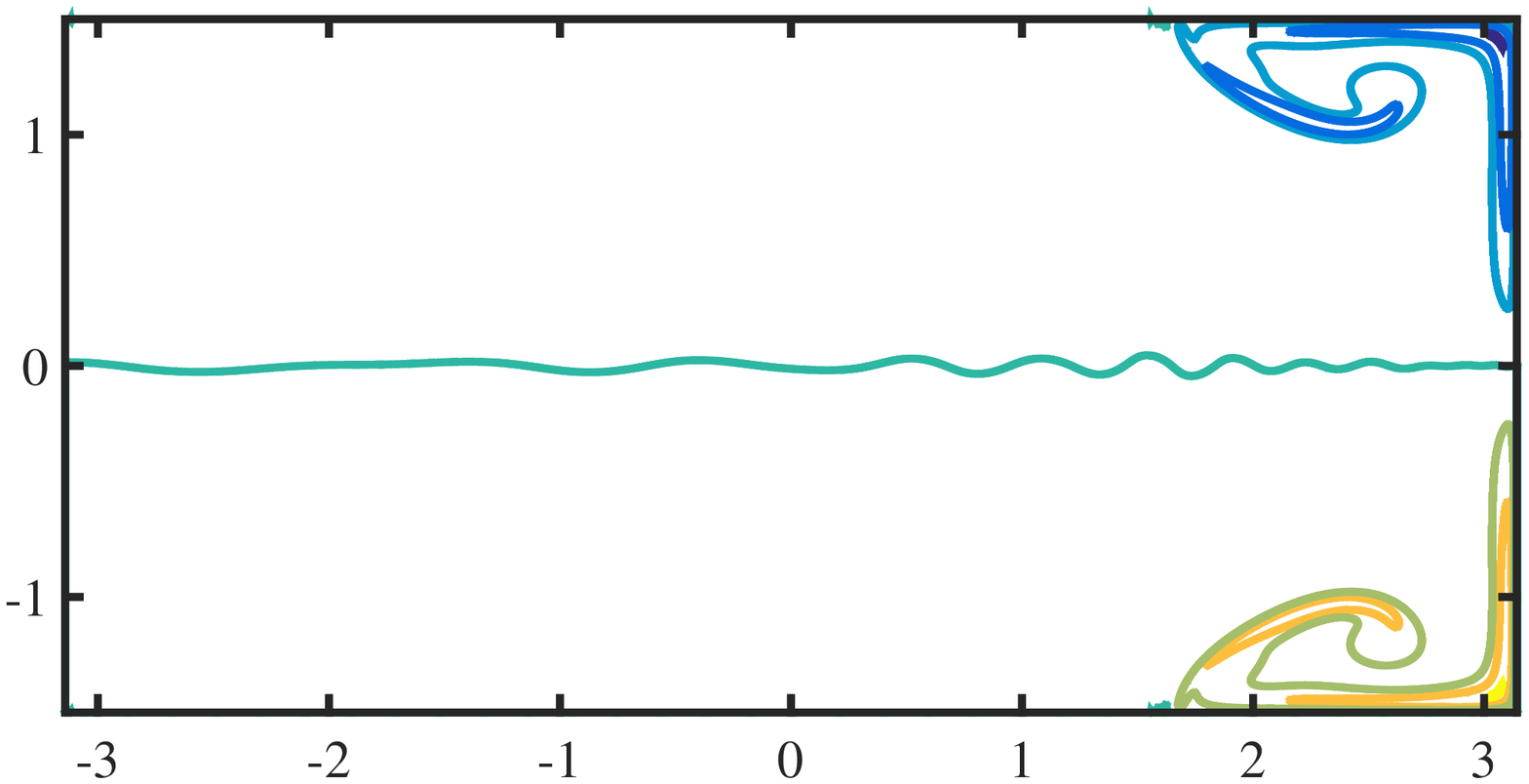}  
  \caption{2D interface, with initially rotational flow, shown at times $t=45$ (left column), $t=60$ (middle column) and $t=75$ (right column). The upper row shows the density perturbation. The less dense fluid is shown in a lighter shade (yellow online) and the denser fluid in a darker shade (blue online). The lower row shows the corresponding vorticity. These calculations used 1600 points and 400 modes in the $x$-direction and 800 points and 200 modes in $y$. The two axes have equal scales.}
   \label{fig3}
  \end{center}
 
\end{figure}

A different 2-D example is shown in Figure \ref{fig3}, with the initial condition being an initial velocity field rather than an interfacial perturbation. Counter-rotating flows have been established at time $t=0$ in the two fluids which now have a horizontal interface at $y=0$. As may be seen in the left frame at $t=45$, a mixing region develops, which is pushed around the outside of the region by the initial flow. In the right frame ($t=75$), it may be seen that an instability is growing on the interface at $y=0$. This instability develops at varying times depending on the number of grid points and modes, and is therefore thought to be initiated by round off error. Once initiated, these disturbances then grow as any Rayleigh-Taylor type disturbance must. This simulation was performed with 1600 by 800 grid points and 400 by 200 modes, and took 32 hours to run.

\subsection{Three-dimensional calculations}
\begin{figure}[!ht]
\includegraphics[width=0.5\columnwidth,clip]{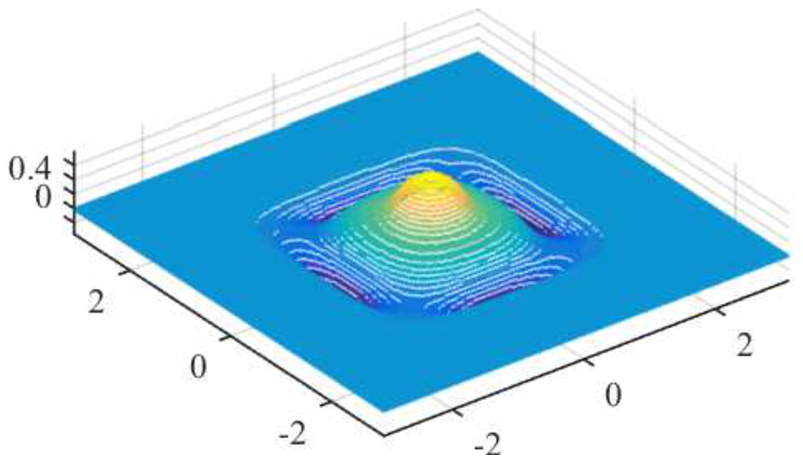}
\includegraphics[width=0.5\columnwidth,clip]{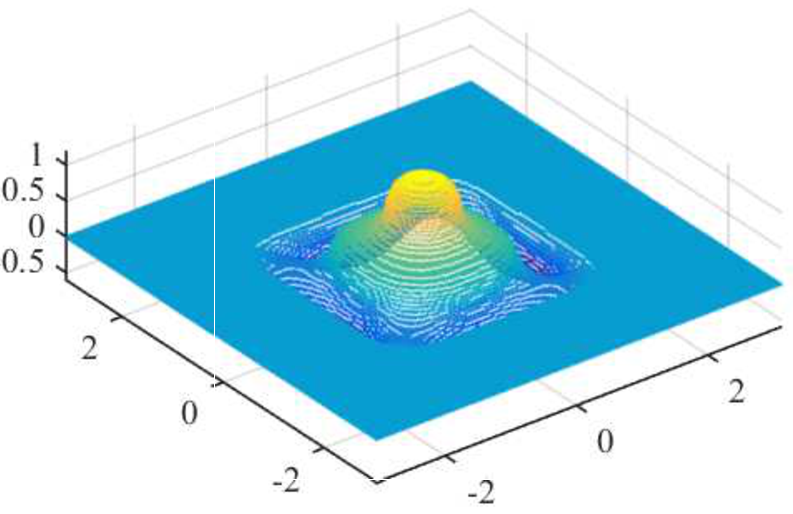}\\
\includegraphics[width=0.5\columnwidth,clip]{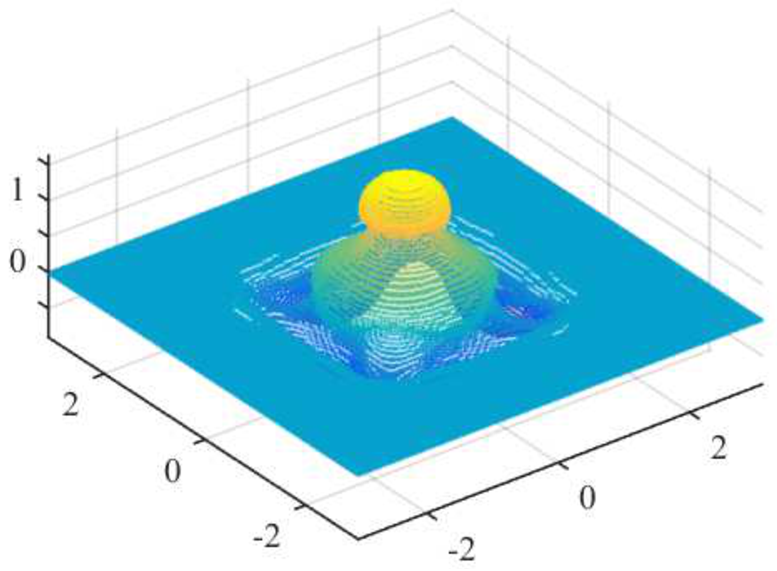}
\includegraphics[width=0.5\columnwidth,clip]{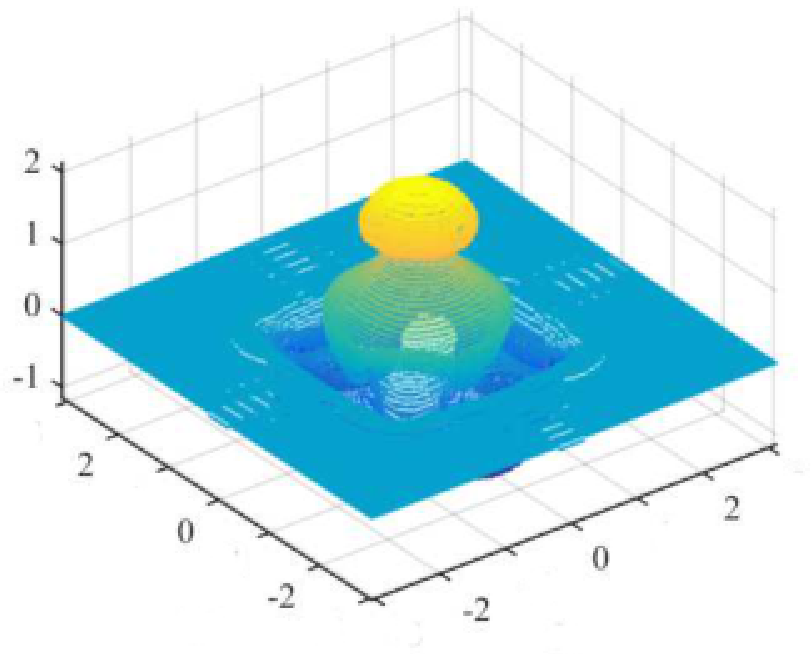}
\caption{3D interface at $t=20$ (upper left), $t=22.5$ (upper right), $t=25$ (lower left) and $t=27.5$ (lower right). The shading (colour online) indicates $z$-value of the interface at that point. These calculations used 192 points in each direction and 36 modes in each variable.}
  \label{fig4}
\end{figure}

A similar procedure was used for generating and displaying the results of the 3D model. As for the 2D system, the sets of spectral coefficients $A_{mnp}, B_{mnp}, C_{mnp}$ were integrated forward in time using the classical fourth order Runge-Kutta method. At any time, the density $\tilde{\rho}$ can be reconstructed from the coefficients $C_{mnp}$. Beginning with an interface with initial shape, symmetric in $x$ and $y$,
\begin{eqnarray}
z_0&=&\frac{\epsilon}{4} (1+\cos(2x))(1+\cos(2y)), -\pi/2<x<\pi/2, -\pi/2<y<\pi/2\nonumber\\
z_0&=&0,\quad \text{elsewhere,}\nonumber
\end{eqnarray}
Similarly to the 2D case shown in Figure \ref{fig2}, the initial interface is approximated by setting $\tilde\rho=0.05/(1+\exp(100(z_0-z)))$. The system is allowed to evolve with $\tilde{\rho}$ reconstructed and plotted at intervals. In order to clarify the display of the interface in three dimensions, only the interfacial points are displayed, rather than colouring each point in the computational region according to its density value, as was done for the 2D model. These interfacial points are defined to be the points where the density perturbation changes from less than $(D-1)/2$ to more than $(D-1)/2$. In Figure \ref{fig4}, we present results for $\epsilon=0.03$, $H_1=1.5, H_2=8$. The viscosity and diffusion parameters were set to $\mu=10^{-4}$ and $\sigma=10^{-4}$. The regular grid of points consisted of $192$ points in each of the $x,y$ and $z$ directions, and $36$ fourier modes for each of $m,n$ and $p$. We have shown the interface at dimensionless times $20, 22.5, 25$ and $27.5$. The upwards movement of the lighter plume, and the downwards movement of the surrounding denser fluid is readily apparent at these times. The plume forms the expected bulbous top which rolls over as it rises. Note that the final frame shows Gibbs' phenomenon \cite{wilb} due to the formation of a very steep density gradient near $x,y=\pm\pi/2$. This can be reduced by the use of greater number of Fourier modes.

\begin{figure}[!ht]
  \includegraphics[width=\columnwidth,clip]{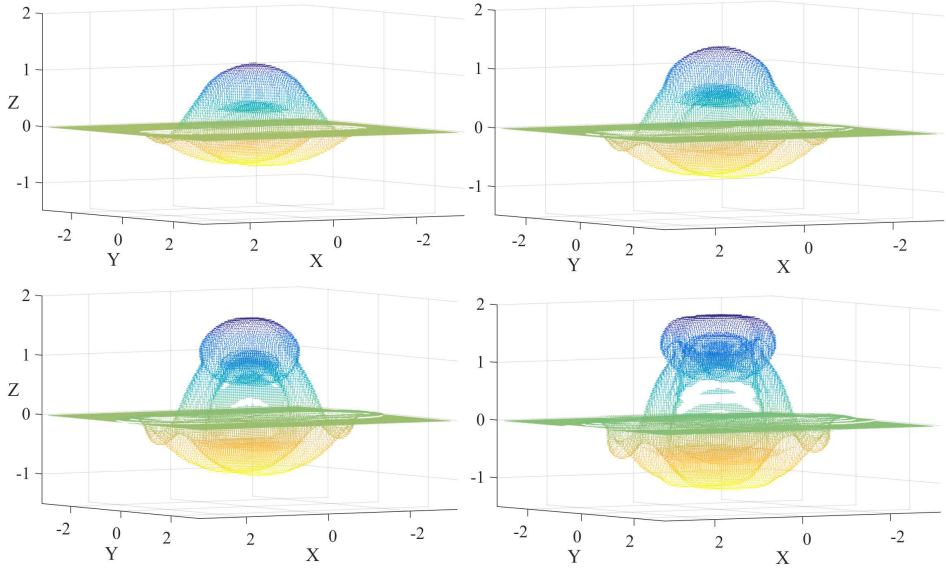}
  \caption{3D interface at $t=22$ (upper left), $t=24$ (upper right), $t=26$ (lower left) and $t=28$ (lower right). As in Figure \ref{fig4}, the colour indicates $z$-value of the interface at that point, but reversed to enhance visualisation of the interface. The interface begins as a small lump, longer in the $x$-direction than in the $y$-direction, before dividing into a structure with two separated `legs'. The flattening effect of a low ceiling at $z=2$ is apparent in the final frame (lower right).}
  \label{fig5}
\end{figure}

As this model has been developed in fully 3D co-ordinates, we are free to give a 3D initial condition, rather than one in which the $x$ components are the same as the $y$ components, as in Figure \ref{fig4}. We thus start with an interface as follows: 
\begin{eqnarray}
z&=&\frac{\epsilon}{4} (1+\cos(4x/3))(1+\cos(4y)), -3\pi/4<x<3\pi/4, -\pi/4<y<\pi/4\nonumber\\
z&=&0,\quad \text{elsewhere.}\nonumber
\end{eqnarray}
Figure \ref{fig5} shows the evolution of this interface for parameters $\epsilon=0.03, H_1=1.5, H_2=2 ,\mu=10^{-3}$ and $\sigma=10^{-4}$. This calculation was performed using $128$ points and $32$ modes in each direction. It can be seen that the interface initially grows exponentially, as expected from the inviscid linear analysis in Section \ref{sec3}. However, the middle section of the interface then splits into a shape with two `legs' connecting the overturning top to the lower part of the interface. In the last frame we see the effect of the low ceiling height in flattening the top of the overturning structure. This calculation took approximately 7 hours.

\begin{figure}[!ht]
\hspace{-1.5cm}\includegraphics[width=1.2\columnwidth,clip]{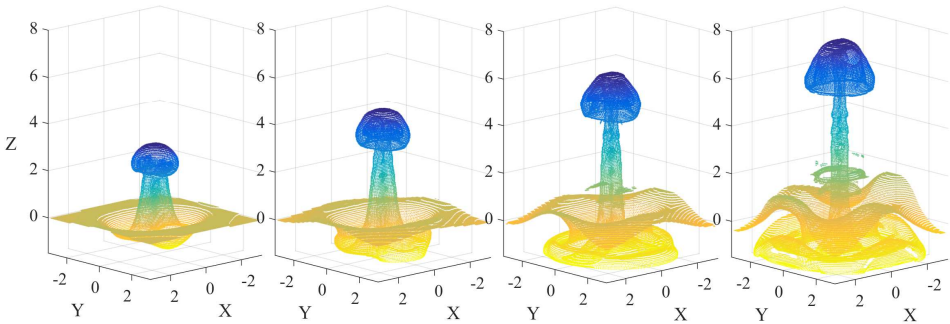}
\caption{3D interface for viscous plume at (from left to right): $t=42$; $t=49.2$; $t=56.4$; and $t=63.6$. The colour indicates $z$-value of the interface at that point. These calculations used 160 grid points and 32 spectral modes in the $x$ ,$y$ and $z$ directions. The axes have equal scales.}
  \label{fig6}
\end{figure}

The initial disturbance shape used to produce Figure \ref{fig5} was again used, but with the Reynolds number decreased by a factor of 10, in order to show the effects of greater viscosity. We have also increased the ceiling height to allow the simulation to run for longer, without being flattened by interaction with the ceiling. The very different results are shown in Figure \ref{fig6}. The much greater viscosity slows the rising of the plume, allowing the formation of a ``mushroom" cap. Towards the end of the run, the stem of the plume is starting to bunch into knots and is shedding rings of fluid near the base. As in Figure \ref{fig4}, we set $H_1=1.5, H_2=8$ and $\sigma=10^{-4}$, but $\mu=10^{-2}$. With 160 grid points and 32 modes in each direction, this simulation took 12 hours.

\begin{figure}
\includegraphics[width=\columnwidth,clip]{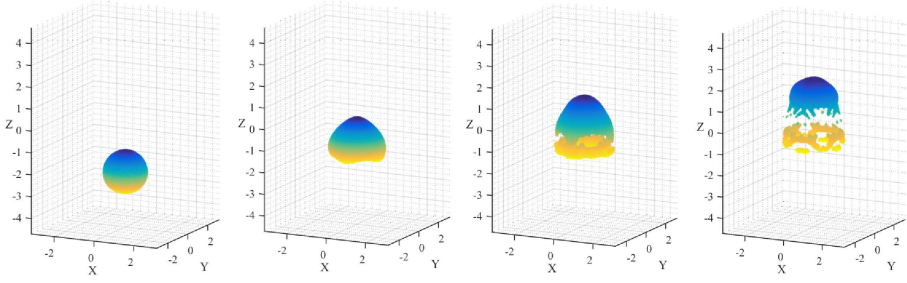}
\caption{3D rising bubble at (from left to right): $t=5$; $t=20$; $t=30$; and $t=40$. The colour indicates $z$-value of the interface at that point. These calculations used 200 grid points and 40 spectral modes in the $x$ and $y$ directions and 300 points and 60 modes in $z$. The axes have equal scales.}
  \label{fig7}
\end{figure}

A very different flow involves an initially spherical drop of fluid rising or falling through a surrounding fluid of slightly different density. In Figure \ref{fig7}, buoyancy is causing a slightly less dense fluid, initially at rest, to rise through the denser surrounding fluid. The initially symmetric sphere (frame 1) can be seen to be distorted by the effect of the walls on the flow of the heavier fluid past the sphere of less dense fluid (frame 2). In the third frame we see that this denser fluid flowing down has caused a ring of material to slow and detach from the bottom of the sphere, leaving a faster rising cap. This simulation was in a box ($-\pi<x<\pi$, $-\pi<y<\pi$, $-3\pi/2<z<3\pi/2$), with an initial sphere of low density fluid of radius 1, centred on the $z$-axis at $z=-0.75\pi$. In this case the density perturbation was simply set as 0 inside the sphere and 0.05 outside. In order to reduce Gibbs' phenomenon due to this sharp interface the initial $C_{mnp}$ coefficients for the density were smoothed using a Lanczos smoothing parameter of 0.05. The run time for this calculation using 200 grid points and 40 spectral modes in the $x$ and $y$ directions and 300 points and 60 modes in $z$, was 50 hours.

\begin{figure}[!ht]
  \includegraphics[width=\columnwidth,clip]{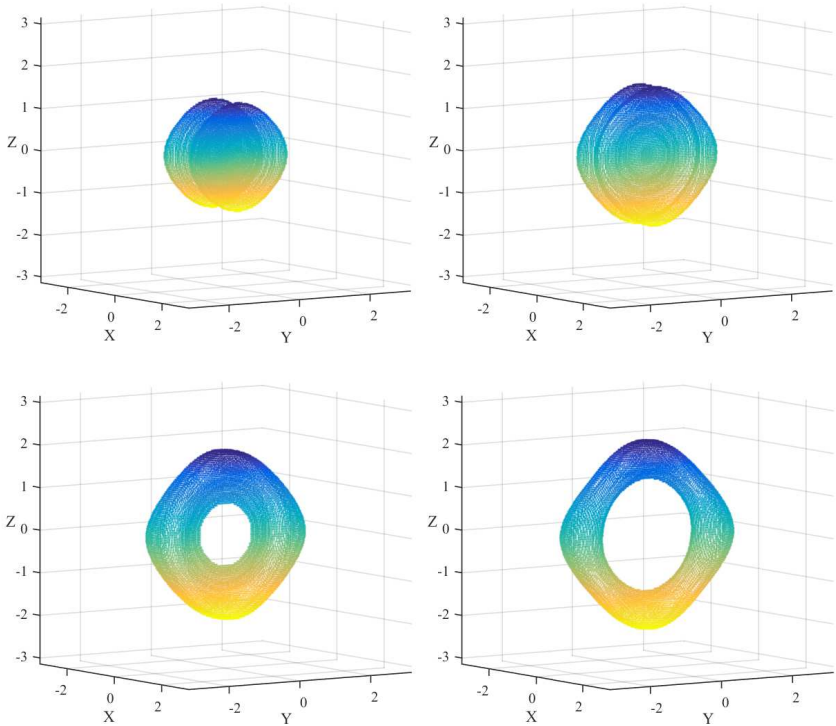}
  \caption{Colliding toroidal blobs at four dimensionless times, $t=120, 240, 360, 480$, from top left to bottom right. The  boundaries are at $\pm\pi$ in each direction. An initial velocity field moves the blobs together until they collide, flatten and spread.}
  \label{fig8}
\end{figure}

A non-Rayleigh-Taylor flow (that is, one with no gravitational effects) is shown in Figure \ref{fig8}, which shows two toroidal blobs of fluid colliding. This simulation was inspired by experimental work \cite{lim} and the fascinating recreation of that experiment \cite{sandlin}. We start with two toroidal blobs of neutral buoyancy ($\tilde{\rho}=0$), with an initial velocity towards the $x=0$ plane. The neutral buoyancy condition is met by setting the gravitational force to zero, $\tilde\rho$ is now just a marker for the different fluid, and may be considered to represent the colour, opacity or some other characteristic of the fluid, which differentiates it from the surrounding medium. The blobs collide and flatten to form a single structure, thinning in $x$ and spreading in $y$ and $z$. The blobs of fluid have an initial surface given by $r^2=(x\pm 1)^2+(0.5-\sqrt{0.9 y^2+z^2})^2$, and an initial velocity field set by $\chi=\pm y/200$ and $\Psi=\pm z/200$ moving the fluid blobs towards the $x=0$ plane. The computational region is bounded by $\pm\pi$ in each direction, with $\sigma=10^{-4}$, $\mu=10^{-3}$. Using 160 points and 40 modes in each direction, this simulation took 47 hours to run. Note that while the fine detail seen in the experimental work of \cite{lim} and \cite{sandlin} is not able to be reproduced at the resolution used here, this simple simulation is included to demonstrate the versatility of the approach presented in this paper.

\subsection{Comparison with linear approximation}

\begin{figure}[!hb]
  \includegraphics[width=\columnwidth,clip]{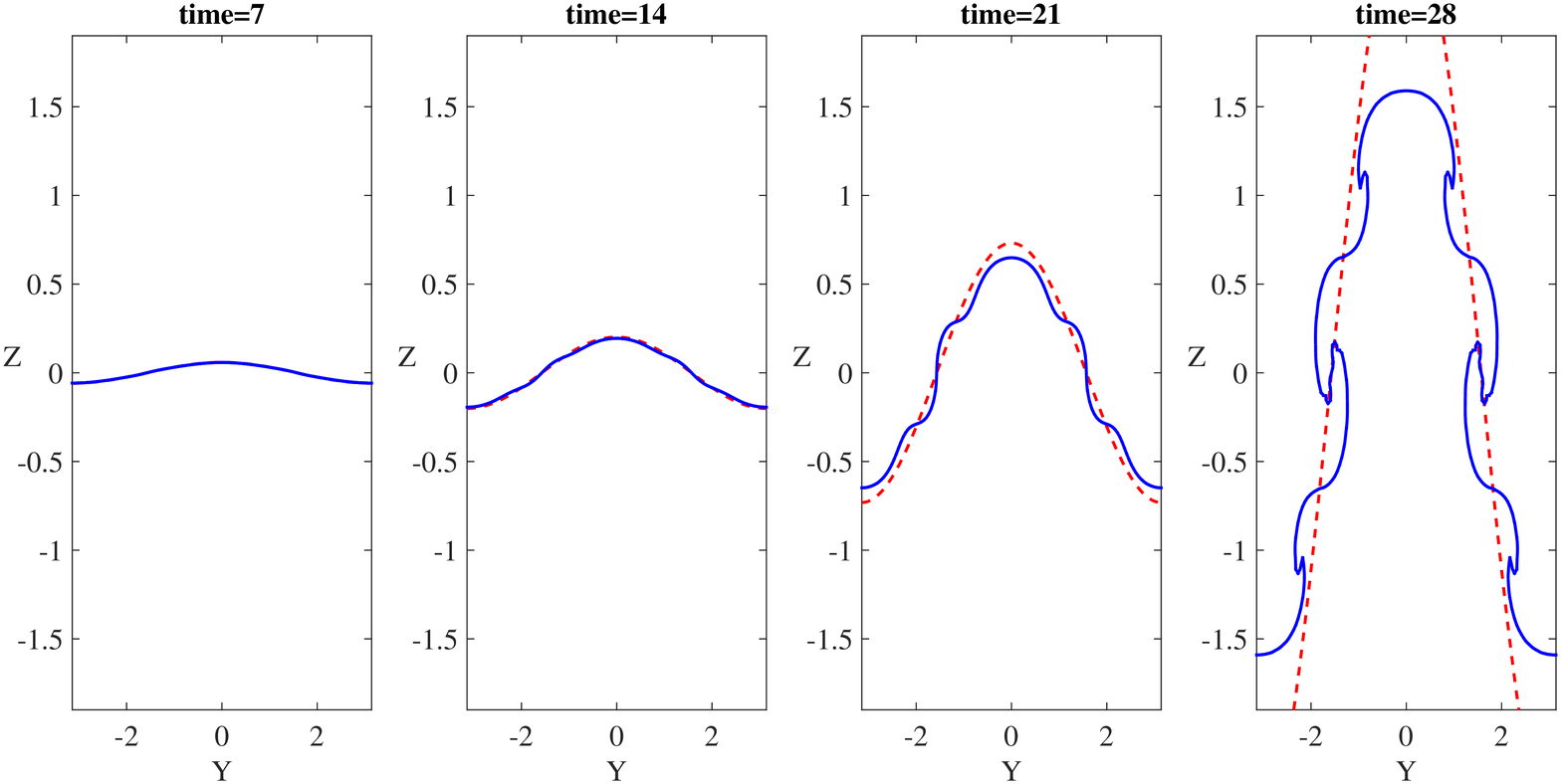}\\
  \includegraphics[width=\columnwidth,clip]{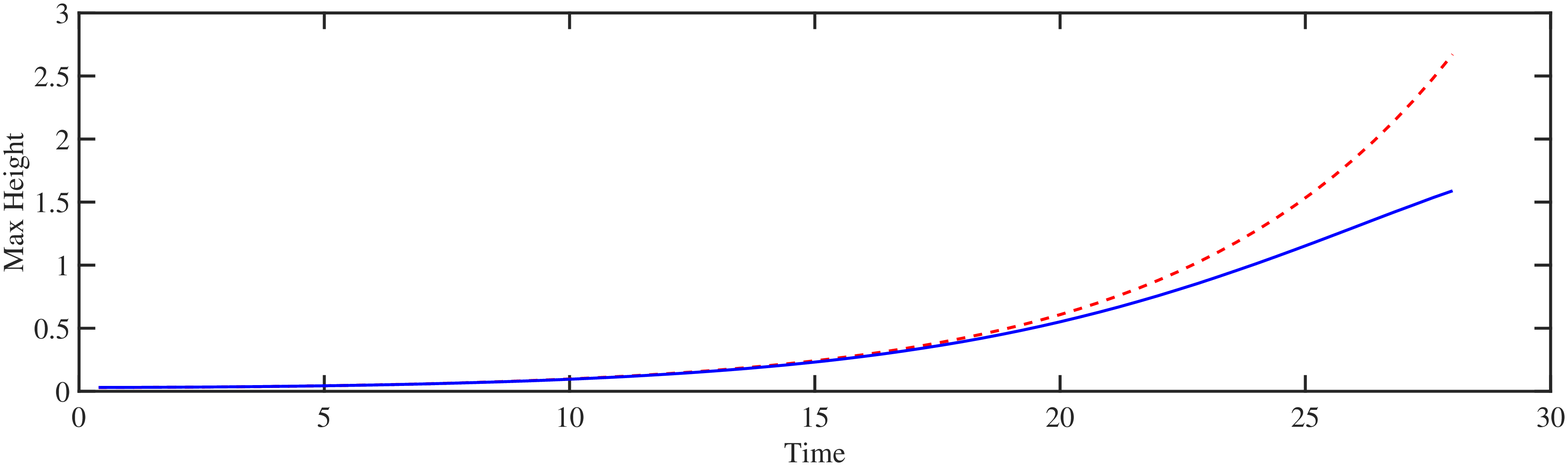}
  \caption{A comparison of the linearised inviscid model (dashed line, red online) with the non-linear viscous numerical model (solid line, blue online) at four dimensionless times, $t=7, 14, 21, 28$. The top and bottom boundaries were set by $H_1=H_2=2$. A slice of each at $x=0$ has been shown in order to compare the linear and non-linear results clearly. The interface in each model starts from rest at $t=0$, with initial profile $z=0.03 \cos x \cos y$. The two models are indistinguishable at $t=7$ and very slightly different at $t=14$. After that time the two models diverge, with the linear model continuing to grow exponentially and the non-linear model slowing and overturning. In the lower panel, we see the height of the interface directly above the origin ($x=y=0$). The linear model (dashed, red) shows exponential growth, while the Boussinesq model (solid, blue) has initial exponential growth, tempered by non-linear effects at later times.}
  \label{fig9}
\end{figure}

The linearised equation (\ref{linear1}) presented in Section \ref{sec3} predicts exponential growth of the interfacial perturbations. This should be very accurate for early times. In the non-linear viscous model, however, at later times viscosity will damp the flow, and the interface will roll over. As a necessary check on the accuracy of the numerical solution which we have provided, it is important to verify that the two approaches do in fact coincide for early times. In Figure \ref{fig9}, we present the linear solution overlaid on the numerical solution, for the case $H_1=H_2=2, D=1.05, \epsilon=0.03,\mu=\sigma=10^{-4}$. The initial interface shape was set as: $z=\epsilon \cos(x)\cos(y)$ for $-\pi<x<\pi$, $-\pi<y<\pi$. The non-linear calculation was performed using $131$ points and $52$ modes in each direction. In order to compare the two clearly, we have taken a central slice at $x=0$. The two models evolve very similarly at first, with the exponential growth predicted by the linear model also dominating the non-linear model. Viscosity and non-linearity become more important factors at later times, damping the exponential growth and causing the interface to roll over, as expected. This agreement with the linearised theory gives confidence that the numerical model is accurately describing the evolution of the flow.

\section{Conclusion}
\label{5}

In this paper we have calculated the evolution of the Rayleigh-Taylor instability in two and three dimensions. By bringing together several modern approaches to the problem, we have accurately modelled the time evolution of a fully 3D fluid system. The use of the spectral method combined with bi-streamfunction approach has provided a very high degree of accuracy, and obviated the need to deal with either the pressure or the incompressibility condition, and allowed the use of exact derivatives and exponential convergence of integrals. This approach has resulted in a very large system of differential equations which can be integrated in parallel, and which is therefore well suited to the capabilities of modern graphics processors. It may be seen that the linearised inviscid model gives excellent agreement with the numerical viscous model at early times, giving us confidence that both the numerical scheme and the linearization are accurate.

While this paper has focussed on the Boussinesq approximation, the triple combination of vorticity equation, bi-streamfunction approach, and spectral method is expected to be directly applicable to other suitable approximations of the Navier-Stokes equations, such as one in which the density ratio is very large (for example in cosmic jets and bubbles). Also, we have shown that this method is able to be used for other interface shapes (the rising sphere in Fig. \ref{fig7}) and for tracking the transport of properties other than density (the colliding blobs of fluid in Fig. \ref{fig8}).
\acks

The authors gratefully acknowledge the financial support for this work provided by Australian Research Council Discovery Grant DP140100094. We are also grateful to two anonymous reviewers, for helpful suggestions on an earlier draft of this paper.

\appendix
\section{Appendix: Differential equations for system of coefficients}
\label{appendix}

In this paper we have presented an approach to solving the viscous Rayleigh-Taylor problem using a spectral method. The set of equations is laid out here for completeness. These are three sets of $m$ by $n$ by $p$ ordinary differential equations which are marched forward in time using the classical fourth order explicit Runge-Kutta method. The integrals were performed using the trapezoidal method, which is exponentially accurate for periodic integrands (\cite{atkinson}, p.288). The first two sets of these equations are derived by substituting the spectral representations (\ref{spectral1}) into the vorticity equation (\ref{i-ns2}) and solving for $\frac{\text{d} A}{\text{d} t}$ and $\frac{\text{d} B}{\text{d} t}$ while the third set is obtained by substituting the series for $\tilde{\rho}$ (\ref{spectral2}) into the transport equation (\ref{transport}).

\begin{eqnarray}
\frac{\text d A_{mnp}}{\text d t}&=&\frac{\alpha_m \alpha_n}{\alpha_p^2 \alpha_{mnp} V \gamma_{mn}}\int_{-\pi}^{\pi}\int_{-\pi}^{\pi}\int_{-h_1}^{h_2}\bigg[(\boldsymbol\zeta \cdot \boldsymbol\nabla)u-(\boldsymbol q \cdot \boldsymbol\nabla)\zeta^X\bigg]C_x S_y S_z\text d z \text d y \text d x\nonumber\\
&+&\frac{\alpha_n^2 + \alpha_p^2}{\alpha_p^2 \alpha_{mnp} V \gamma_{mn}}\int_{-\pi}^{\pi}\int_{-\pi}^{\pi}\int_{-h_1}^{h_2}\bigg[(\boldsymbol\zeta \cdot \boldsymbol\nabla)v-(\boldsymbol q \cdot \boldsymbol\nabla)\zeta^Y\bigg]S_x C_y S_z\text d z \text d y \text d x \nonumber\\
&-&\frac{\alpha_m}{\alpha_{mnp}}C_{mnp}-\mu\alpha_{mnp}A_{mnp} \nonumber\\
\frac{\text d B_{mnp}}{\text d t}&=&\frac{\alpha_m^2 + \alpha_p^2}{\alpha_p^2 \alpha_{mnp} V \gamma_{mn}}\int_{-\pi}^{\pi}\int_{-\pi}^{\pi}\int_{-h_1}^{h_2}\bigg[(\boldsymbol\zeta \cdot \boldsymbol\nabla)u-(\boldsymbol q \cdot \boldsymbol\nabla)\zeta^X\bigg]C_x S_y S_z\text d z \text d y \text d x\nonumber\\
&+&\frac{\alpha_m \alpha_n}{\alpha_p^2 \alpha_{mnp} V \gamma_{mn}}\int_{-\pi}^{\pi}\int_{-\pi}^{\pi}\int_{-h_1}^{h_2}\bigg[(\boldsymbol\zeta \cdot \boldsymbol\nabla)v-(\boldsymbol q \cdot \boldsymbol\nabla)\zeta^Y\bigg]S_x C_y S_z\text d z \text d y \text d x \nonumber\\
&-&\frac{\alpha_n}{\alpha_{mnp}}C_{mnp}-\mu\alpha_{mnp}B_{mnp} \nonumber\\
\frac{\text d C_{mnp}}{\text d t}&=&-\frac{1}{V \gamma_{mn}}\int_{-\pi}^{\pi}\int_{-\pi}^{\pi}\int_{-h_1}^{h_2}\bigg[(\boldsymbol q \cdot \boldsymbol\nabla)\tilde{\rho}\bigg]C_x C_y S_z\text d z \text d y \text d x-\sigma \alpha_{mnp}C_{mnp}
\label{fulleqns}
\end{eqnarray}

In these expressions, the quantities $\gamma_{mn}$ are constants that arise in Fourier analysis for the $(m,n)$-th mode at a fixed value of $p$, For $m,n$ both non-zero, $\gamma_{mn}=1$, but $\gamma_{01}=\gamma_{10}=2$ and $\gamma_{00}=4$. The volume constant $V=4\pi^2(H_1+H_2)/8$ also results from Fourier decomposition and the orthogonality of the trigonometric modes. Additionally, the following abbreviations have been used:
\begin{eqnarray}
S_x &\equiv& \sin(\alpha_m x) \nonumber\\
C_x &\equiv& \cos(\alpha_m x) \nonumber\\
S_y &\equiv& \sin(\alpha_n y) \nonumber\\
C_y &\equiv& \cos(\alpha_n y) \nonumber\\
S_z &\equiv& \sin(\alpha_p (z+H_1)) \nonumber\\
C_z &\equiv& \cos(\alpha_p (z+H_1)). \nonumber
\end{eqnarray}

We finally note that if half-mode basis functions are used (for example, $\alpha_m=m/2$) as discussed in the last paragraph of Section \ref{sec4}, some small modifications are required to the equations (\ref{fulleqns}) as not all of the orthogonality conditions hold in that case.

\bibliographystyle{unsrtnumbered}

\bibliography{test}

\end{document}